\documentclass[a4,traditabstract]{aa}

\usepackage{graphicx,amssymb,amsmath}
\usepackage{txfonts}
\usepackage{natbib}

\newcommand{\ain}{a_{\rm in}}
\newcommand{\aout}{a_{\rm out}}
\newcommand{\elik}{{\mathbf K}}

\newcommand{\so}{\Sigma_0}
\newcommand{\st}{\Sigma_{\rm tot}}
\newcommand{\psit}{\Psi^{\rm thin}}
\newcommand{\psif}{\Psi^{\rm flat}}

\newcommand{\ks}{k_{\rm s}}
\newcommand{\ms}{m_{\rm s}}

\def\atan{{\rm arctan \,}}

\begin{document}

\title{A local prescription for the softening length\\ in self-gravitating gaseous discs}


\author{Jean-Marc Hur\'e\inst{1,2} \and Arnaud Pierens\inst{3}}

\offprints{jean-marc.hure@obs.u-bordeaux1.fr}

\institute{Universit\'e de Bordeaux, OASU
\and
CNRS/INSU-UMR 5804/LAB; BP 89, 33271 Floirac cedex, France
\and
LAL-IMCCE/USTL, 1 Impasse de l'Observatoire, F-59000 Lille, France
}

\date{Received ??? / Accepted ???}

\abstract
{In 2D-simulations of self-gravitating gaseous discs, the potential is often computed in the framework of "softened gravity" initially designed for $N$-body codes. In this special context, the role of the softening length $\lambda$ is twofold: i) to avoid numerical singularities in the integral representation of the potential (i.e., arising when the separation $|\vec{r} -\vec{r}'| \rightarrow 0$), and ii) to acount for stratification of matter in the direction perpendicular to the disc mid-plane. So far, most studies have considered $\lambda$ as a free parameter and various values or formulae have been proposed without much mathematical justification. In this paper, we demonstrate by means of a rigorous calculus that it is possible to define $\lambda$ such that the gravitational potential of a flat disc coincides at order zero with that of a geometically thin disc of the same surface density. Our prescription for $\lambda$, valid in the local, axisymmetric limit, has the required properties i) and ii). It is mainly an analytical function of the radius and disc thickness, and is sensitive to the vertical stratification. For mass density profiles considered (namely, profiles expandable over even powers of the altitude), we find that $\lambda$ : i) is independant of the numerical mesh, ii) is always a fraction of the local thickness $H$, iii) goes through a minimum at the singularity (i.e., at null separation), and iv) is such that $0.13 \lesssim \lambda/H \lesssim 0.29$ typically (depending on the separation and on density profile). These results should help us to improve the quality of 2D- and 3D-simulations of gaseous discs in several respects (physical realism, accuracy, and computing time).}


\maketitle

\section{Introduction}
\label{sec:intro}

The computation of gravitational potentials and forces is a critical step in many astrophysical problems where gravity plays an important role. Because Newton's law of gravitation diverges as the relative separation $|\vec{r}_i-\vec{r}_j|$ between particles vanishes, one classically adds a positive constant $\lambda$ referred as the ``softening length''. This technique of singularity avoidance, initially developped to prevent binary collisions and particle evaporation in $N$-body simulations, has led to the concept of ``softened gravity'' \citep{he88,anca88,sommer98,nelson06}. Soon, people realized that the use of a softening length introduces a noticeable bias in models, especially for the stability properties of stellar systems, and there have been several attempts to search for the most appropriate $\lambda$-value \citep[e.g.,][]{romeo94,romeo97,sommer98,dehnen01}.

Softened gravity has also been employed in the computation of the gravitational potential of {\it gaseous discs} \citep[e.g.,][]{paplin89,ars89,sayo90,shu90,mosa94,sterzik95,lka97,lka98,tremaine01,caunt01,baruteaumasset08,li09} with the additional justification that the softening length takes into account the vertical stratification of matter. Various prescriptions, generally in the form of a function of the cylindrical radius and/or of the disc parameters, have been adopted (see references above) without convergence towards a ``universal formula''. As for systems of particles, it is recognized that the softening length can dramatically affect the stability of gaseous discs (again, see references hereabove). It is therefore fundamental to ask i) whether or not softened gravity can definitively help to determine or mimic the {\it Newtonian potential of a gaseous disc} with a certain level of accuracy, and ii) eventually which formula | if one exists | is the most appropriate. This is the aim of this paper. By equating the softened potential of a flat disc to that of a geometrically thin disc, we find that we can define a softening length $\lambda$ which has the required properties. We therefore report the first reliable formula for the softening length on the basis of approximate but rigorous calculus. In the axisymmetric limit, $\lambda$ is found to be a sharp function of the relative separation $|\vec{r} -\vec{r}'|$; it is a fraction (of the order of $\frac{1}{2e}$ at the singularity) of the disc local thickness \citep[see also][]{hp06,baruteaumasset08}, and it does not depend on the numerical resolution \citep[see][]{li09}. The formula, easy to implement into hydrodynamical codes of self-gravitating 2D- and 3D-discs, will enable to increase the degree of reaslim of simulations, both by {\it preserving the Newtonian character} of the potential and force field, and by {\it accounting for the vertical structure}.

The paper is organized as follows. In Sect. \ref{sec:midplanepot}, we recall the expression for the midplane gravitational potential of a geometrically thin disc as well as that of a flat (i.e., zero thickness) disc. Because of the hypothesis of axial symmetry, the integral kernel contains a complete elliptic integral of the first kind that can be appropriately expanded at the singularity $|\vec{r} -\vec{r}'|=0$ and its neighborhood. In Sect. \ref{sec:chif}, we estimate the effect of vertical stratification by performing the analytical integration along the $z$-direction. Various mass density profiles corresponding to finite size discs are considered, namely the homogeneous case and mixtures of even powers of the altitude. In Sect. \ref{sec:sg}, we show that this calculus naturally leads to a local prescription for the softening length $\lambda$. We present a table showing the great diversity (and incoherence) of prescriptions used so far and derive the general relation for $\lambda$. We discuss the sensitivity of the softening length to the separation and disc aspect ratio. In Sect. \ref{sec:conclusion}, we summarize the results and suggest a few possible extensions and generalizations of this work.

\section{Mid-plane gravitational potential in both flat discs and geometrically thin discs : local treatment.}
\label{sec:midplanepot}

\subsection{Notation and background}

\begin{figure}[h]
\includegraphics[width=8.9cm]{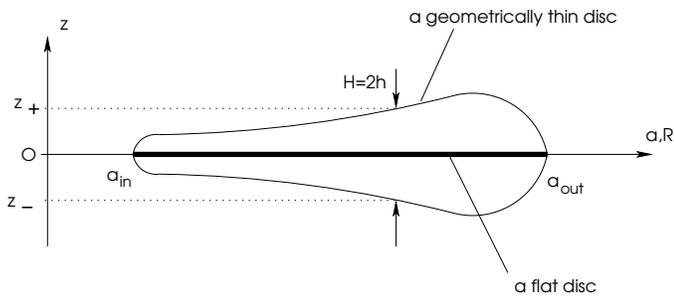}
\caption{A flat disc and a geometrically thin disc, axially symmetric and symmetric with respect to the mid-plane. The disc edges are $\ain$ and $\aout$, $h$ is the semi-thickness, $\st$ is the total surface density, and $a$ and $R$ are cylindrical radii.}
\label{fig:sys.xfig.eps}
\end{figure}

We consider two axially symmetric discs  (see Fig. \ref{fig:sys.xfig.eps}) : i) a flat (i.e., zero thickness) disc with inner edge $\ain$, outer edge $\aout$, and total surface density $\st$; and ii) a geometrically thin disc with the same edges and same total surface density, but local thickness $H = 2h>0$.
 For the geometrically thin disc, the condition
\begin{equation}
\left(\frac{H}{a}\right)^2 \ll 1
\label{eq:gtda}
\end{equation}
 is assumed at any radius $a \in [\ain,\aout]$ \citep[][]{pringle81}. Moreover, we have
\begin{equation}
\int_{z_-}^{z+}{\rho(z) dz} = \st,
\label{eq:st}
\end{equation}
where $\rho$ is the mass density at the altitude $z$ from the mid-plane, $z_-$ is the altitude of the bottom of the disc, and $z_+~=z_-+2h$ is for the top. In general, $\rho$, $\st$, $z_-$, and $h$ are expected to depend on the radius $a$. In the following, we shall also consider symmetry with respect to the mid-plane, which is the case in most disc models (not true for warped discs for instance). This means that $z_-+z_+=0$, and $z_+=h$. For $z+ \rightarrow 0$ for all $a$, the two discs therefore become  equivalent.

For the thin disc, the mid-plane gravitational potential at radius $R$ is given by the double integral \citep[e.g.][]{durand64}
\begin{equation}
\psit(R) = -2G\int_{\ain}^{\aout}{\sqrt{\frac{a}{R}}  da\; \int_{-h}^{h}{\rho k \elik(k) dz}},
\label{eq:psit}
\end{equation}
where 
\begin{equation}
k = \frac{2\sqrt{aR}}{\sqrt{(a+R)^2+z^2}}
\label{eq:kz}
\end{equation}
is the modulus and $\elik$ is the complete elliptic integral of the first kind (Legendre form). For the flat disc, we have $\rho(z) = \st \delta(z)$ everywhere. Consequently, the potential at radius $R$ is
\begin{equation}
\psif(R)=-2G \int_{\ain}^{\aout}{\sqrt{\frac{a}{R}}\; \st m \elik(m) da},
\label{eq:psif}
\end{equation}
where
\begin{equation}
m=\frac{2\sqrt{aR}}{a+R}
\end{equation}
is the associated modulus (note that $k=m$ for $z=0$). Both Eqs. \ref{eq:psit} and \ref{eq:psif} involve a logarithmic singularity when the modulus of the elliptic integral approaches unity. In spite of this divergence, the potential is generally finite at any radius. It is interesting to see that Eq. \ref{eq:psit} can also be written as
\begin{equation}
\psit(R) = -2G\int_{\ain}^{\aout}{\sqrt{\frac{a}{R}} \st \chi da},
\label{eq:psitbis}
\end{equation}
where
\begin{equation}
\st \chi = \int_{-h}^h{\rho  k \elik(k) dz}.
\label{eq:chi}
\end{equation}
Clearly, $\chi$ is a function of $a$, $R$, and $h$ and contains the effects of vertical stratification. So, if $\chi$ is  known preliminarily at all radii and for a given disc structure (edges, thickness, mass density profile), $\psit$ is found {\it from a single integral over the radius} (as for $\psif$). It is therefore advantageous to have a {\it one-dimensional formula that describes a bi-dimensional distribution of matter} (in the axially symmetric case), especially in terms of computing time. By comparing Eqs. \ref{eq:psif} and \ref{eq:psitbis}, it is tempting to set $\chi = \ms \elik(\ms)$, where $\ms$ is a certain modulus to be determined (see Sect. \ref{sec:sg}).

\subsection{Expansion around the singularity}

The presence of the function $\elik$ does not allow us to derive $\chi$ analytically for any mass density distribution. The expansion of $\elik(k)$ appropriate for treating the logarithmic singularity is \citep[e.g.][]{gradryz65}
\begin{equation}
\elik(k)= \sum_{n=0}^{\infty}{\beta_n P_n(k') {k'}^{2n}},
\label{eq:kseries}
\end{equation}
where $k' = \sqrt{1-k^2}$ is the complementary modulus, and 
\begin{equation}
\begin{cases}
P_0= \ln \frac{4}{k'},\\\\
P_{n+1}(k') = P_n(k') - \frac{1}{(2n+1)(n+1)}, \qquad \text{for } n \ge 0,\\\\
\beta_0=1,\\\\
\beta_{n+1} = \left[\frac{(2n+1)!!}{2(n+1)!!}\right]^2,\qquad \text{for } n \ge 0.
\end{cases}
\label{eq:gamman}
\end{equation}
By construction, this expansion is efficient when $k' \rightarrow 0$, which corresponds to $k \rightarrow 1$. It can be shown that the condition ${k'}^2 \ll 1$ implies that
\begin{equation}
\left(\frac{h}{a+R}\right)^2 \ll 1,
\end{equation}
and
\begin{equation}
\frac{|R-a|}{h} \ll \frac{a}{h}.
\end{equation}
The first inequality is automatically fulfilled within the geometrically thin disc approximation (see Eq. \ref{eq:gtda}). The second inequality means that the present calculus is valid only {\it at the singularity $R=a$} and {\it for a few disc thicknesses in radius}. This is precisely the radial domain where the numerical determination of $\psit$ is tricky \citep[e.g.,][]{syc90}, because of the divergence of $\elik(k)$. We note that there is no special constraint on the local shape $h(a)$ of the disc provided that this remains geometrically thin.
 
To the lowest order, $\elik(k) \approx \ln\frac{4}{k'}$ and $k \approx 1$, and Eq. \ref{eq:chi} becomes :
\begin{equation}
\st \chi \approx 2 \int_0^{h}{\rho \ln \frac{4}{k'}dz},
\label{eq:chiorder1}
\end{equation}
with an error of ${\cal O}({k'}^2 \ln k')$.

\section{Effect of vertical stratification : an estimate of the $\chi$-function}
\label{sec:chif}

\subsection{Vertically homogeneous discs}

\begin{figure}
\includegraphics[width=8.9cm, bb=44 51 706 529, clip=]{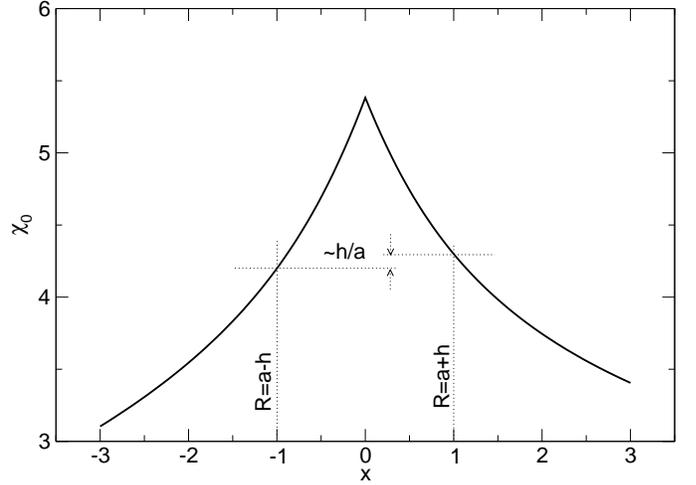}
\caption{Variation of $\chi_0$ with $x$ according to Eq. \ref{eq:chispec0}.}
\label{fig:chi0.eps}
\end{figure}

We can estimate $\chi$ from Eq.\ref{eq:chiorder1} in a few particular cases, for instance when the disc is vertically homogeneous ($\chi$ is denoted $\chi_0$ in this case). If $\rho$ does not depend on $z$, then $\st =2 \rho h \equiv \so$ and we find that 
\begin{equation}
\chi_0 = \ln 4 - \ln k'_\pm-\ln f_0,
\label{eq:chispec0}
\end{equation}
where
\begin{equation}
\begin{cases}
\label{eq:chispec2}
{k'_\pm}^2 = \frac{(a-R)^2+h^2}{(a+R)^2+h^2},\\\\
\eta_\pm = \frac{h}{a \pm R}\\\\
\ln f_0 =  \frac{1}{\eta_-}  \atan \eta_-  - \frac{1}{\eta_+} \atan \eta_+.
\end{cases}
\end{equation}
This is a complicated function of $R/a,$ where the local aspect ratio $2h/a$ is the unique parameter. If we define the variable
\begin{equation}
x=\frac{R-a}{h}
\end{equation}
which measures the radial separation from the singularity $R=a$ in units of the disc semi-thickness $h$, and
\begin{equation}
\epsilon =\frac{h}{2a},
\end{equation}
which is the quarter of the aspect ratio, we find that
\begin{flalign}
\nonumber
\chi_0 = \ln 4 &- \ln \sqrt{\frac{(1+x^2)(1+\epsilon x)^2}{\epsilon^2+(1+\epsilon x)^2}}\\
&-x \, \atan \frac{1}{x} + \frac{1+ \epsilon x}{\epsilon} \, \atan \frac{\epsilon}{1+ \epsilon x}.
\label{eq:chispec0_explicit}
\end{flalign}
The function $\chi(x)$ is plotted in Fig. \ref{fig:chi0.eps} for $h/a=0.1$ (that is $\epsilon=0.05$). To order zero around $x=0$, Eq. \ref{eq:chispec0} reads
\begin{equation}
\chi_0 \approx 1 +  \ln \frac{4}{\epsilon} + \epsilon x - \frac{\pi}{2} |x| - \ln \sqrt{1+x^2} + x \, \atan x.
\label{eq:chi0_approx}
\end{equation}
We see that $\chi_0$ reaches a maximum value of $1 + \ln \frac{4}{\epsilon} \gg 1$ at $R=a$, and is slightly asymmetric with respect to $x=0$, since we have $\chi_0(x=+1) \approx \chi_0(x=-1)+ 2\epsilon$.

\subsection{Effect of vertical stratification}

We can probe the sensitivity of $\chi$ to vertical stratification by considering the following profiles :
\begin{equation}
\rho(z)= 
\begin{cases}
\rho_0 \left[1-\left(\frac{z}{h}\right)^{2q}\right] & \text{for } |z| \le h,\\ \\
0 & \text{otherwise},
\label{eq:vertprofile}
\end{cases}
\end{equation}  
where $\rho_0$ is the density at the disc mid-plane (possibly a function of $a$) and $q \ge 1$ is an integer. These correspond to finite size discs. When $q=1$, the profile is close to a Gaussian distribution, a case found for instance in vertically isothermal disc at hydrostatic equilibrium \citep[e.g.,][]{chiang97,hirose06,edgar08}. As $q$ increases, the vertical profile becomes flatter and flatter. For $q \rightarrow \infty$, we recover the homogeneous case considered above.

We can calculate $\chi$ again from Eq. \ref{eq:chiorder1}, but using Eq.~\ref{eq:vertprofile}. From Eq. \ref{eq:st}, we get $\st = \frac{2q}{2q+1} \so$, and then
\begin{flalign}
\chi 
     & = \chi_0 + \frac{1}{2q} \left( \chi_0 - \chi_q \right),
\label{eq:chi_mix}
\end{flalign}
where we have defined
\begin{equation}
\chi_q = (2q+1) \int_0^{1}{\left(\frac{z}{h}\right)^{2q} k \elik(k) d \frac{z}{h}}.
\label{eq:chiq}
\end{equation}
We can perform the integration using the expansion of $\elik(k)$ around $k'=0$, as above. To the lowest-order, we have
\begin{equation}
\chi_q = \ln 4 - \ln k'_\pm - \ln f_q,
\label{eq:chispecn}
\end{equation}
where
\begin{flalign}
\label{eq:lnfn}
\ln f_q & = (-1)^q \left[ \frac{\atan \eta_-}{\eta_-^{2q+1}} - \frac{\atan \eta_+}{\eta_+^{2q+1}} \right.\\
& \qquad \qquad \qquad   \left. -\sum_{k=0}^q{(-1)^k  \frac{\eta_-^{2(k-q)}-\eta_+^{2(k-q)}}{2k+1}} \right].
\nonumber
\end{flalign}

\begin{figure}
\includegraphics[width=8.9cm, bb=44 51 706 529, clip=]{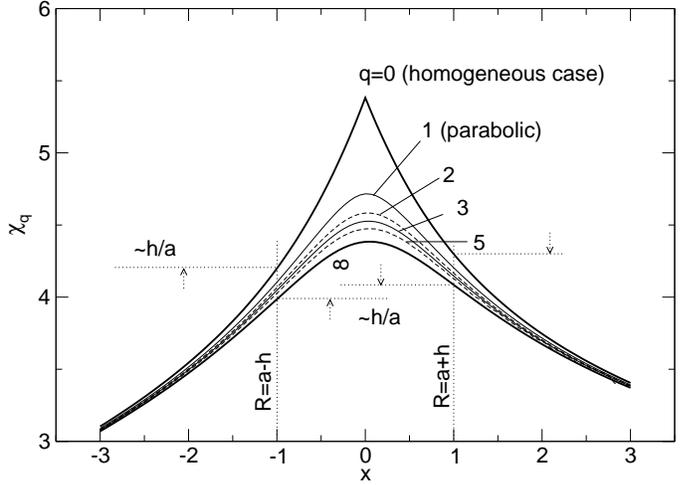}
\caption{Variation of $\chi_q$ with $x$ according to Eq. \ref{eq:chispecn} for $h/a=~0.1$ and different values of $q$, namely $q=0$ (the homogeneous case), $q=1$ (the parabolic case), $q=2, 3, 5$ and $q \rightarrow \infty$.}
\label{fig:chiq.eps}
\end{figure}

This expression can be rewritten in a different form since each sum represent the first $q+1$ terms of the expansion of the $\atan$ function. In particular, a form appropriate to numerical computations around the singularity is
\begin{flalign}
\ln f_q & = (-1)^q \left[ \eta_-^{-2q} \left(  \frac{\pi}{2|\eta_-|} - \frac{1}{\eta_-} \atan \frac{1}{\eta_-} \right)  \right.\\
& \qquad \left. -\sum_{k=0}^q{\frac{(-1)^k}{2k+1} \eta_-^{2(k-q)}} \right] + \eta_+^2\sum_{k=0}^\infty{\frac{(-1)^k \eta_+^{2k}}{2(k+q)+3} }.
\nonumber
\end{flalign}
We see that $\chi_q$ is a function of $R/a$, and $h/a$ is the parameter. Using again the variable $x$ and the $\epsilon$-parameter, we find
\begin{flalign}
\label{eq:chispecq_explicit}
\chi_q = \ln 4 &- \ln \sqrt{\frac{(1+x^2)(1+\epsilon x)^2}{\epsilon^2+(1+\epsilon x)^2}}\\
\nonumber
& - (-1)^q \left[ x^{2q} \left(  \frac{\pi}{2}|x| - x \, \atan x \right)  -\sum_{k=0}^q{\frac{(-1)^k}{2k+1} x^{2(q-k)}} \right] \\
\nonumber
& \qquad  - \left(\frac{\epsilon}{1+\epsilon x}\right)^2\sum_{k=0}^\infty{\frac{(-1)^k }{2(k+q)+3} \left(\frac{\epsilon}{1+\epsilon x}\right)^{2k}}.
\end{flalign}
The function $\chi_q$ is displayed versus $x$ in Fig. \ref{fig:chiq.eps} for $h/a=0.1$ and different values of $q$. Around $x=0$, Eq. \ref{eq:chispecq_explicit} becomes
\begin{flalign}
\nonumber
\chi_q & \approx \frac{1}{2q+1} + \ln \frac{4}{\epsilon} + \epsilon x - \ln \sqrt{1+x^2} \\
& \qquad \qquad - (-1)^q x^{2q} \left(\frac{\pi}{2}|x| - x \, \atan x \right),
\label{eq:chiq_approx}
\end{flalign}
and we note that
\begin{equation}
\chi_0 - \chi_q = \frac{2q}{2q+1} + \left[(-1)^q x^{2q} -1 \right] \left( \frac{\pi}{2}|x| - x\,\atan x \right).
\end{equation}
The maximum of $\chi_q$ still occurs at $R=a$ (see Fig. \ref{fig:chiq.eps}) :
\begin{equation}
\chi_{q, \rm max} \approx \chi_{0, \rm max} - \frac{2q}{2q+1}.
\end{equation}
Again, we note that the difference $\chi_q(R=a+h) - \chi_q(R=a-h) \approx 2 \epsilon$ is insensitive to $q$ at the actual order of precision.

\begin{figure}
\includegraphics[width=8.9cm, bb=14 51 706 529, clip=]{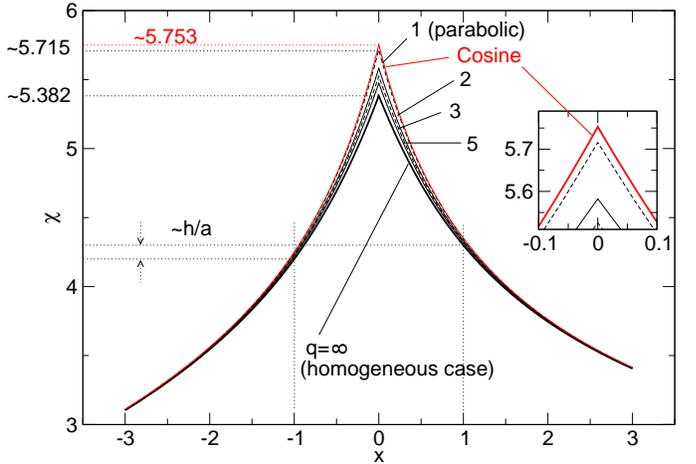}
\caption{$\chi$ versus $x$ for the mass density profile defined by Eq. \ref{eq:vertprofile}. The cosine profile discussed in Sect \ref{subsec:lambda} is also shown.}
\label{fig:chi.eps}
\end{figure}

Finally, for the vertical profile defined by Eq. \ref{eq:vertprofile}, $\chi$ follows from Eqs. \ref{eq:chispec0_explicit}, \ref{eq:chi_mix}, and \ref{eq:chispecq_explicit}. We see that $\chi$ reaches a maximum at $R=a$, and
\begin{equation}
\chi_{\rm max} \approx 1 + \ln \frac{4}{\epsilon} + \frac{1}{2q+1},
\end{equation}
for $q \ge 1$. For large values of the $q$-parameter, $\chi \rightarrow \chi_0$. Figure \ref{fig:chi.eps} displays $\chi$ versus $x$ for $h/a=0.1$ and different values of $q$. It is remarkable that $\chi$ is very weakly sensitive to the vertical profile and that $\chi$ remains very close to $\chi_0$ (the homogeneous case), especially for $|x| \gtrsim 1$.

\subsection{Full generalization}

The generalization of the above result is relatively straightforward for profiles of the form
\begin{equation}
\rho(z)= \rho_0 \sum_{n=0}^\infty{c_n \left(\frac{z}{h}\right)^{2n}},
\label{eq:genprofile}
\end{equation}  
where $c_n$ are real coefficients. In this case, we find that
\begin{equation}
\chi = \frac{1}{\st}\sum_{n=0}^\infty{\frac{c_n \chi_n}{2n+1}}, \quad \text{with} \; \st = \so \sum_{n=0}^\infty{\frac{c_n}{2n+1}},
\label{eq:gen_chi}
\end{equation}
where $\chi_n$ is found from Eq. \ref{eq:chispecq_explicit}. Again, $\chi$ is a function of $x$, and $\epsilon$ is the parameter. It is especially convenient to write $\chi$ as
\begin{equation}
\chi = \chi_0 + \delta \chi_0,
\end{equation}
where $\delta \chi_0$ is the deviation relative to the homogeneous case. We note that $\delta \chi_0$ is generally small compared to $\chi_0$ as long as the bulk of the mass is localized in the disc mid-plane. We conclude that {\it $\chi$ is determined mainly by the homogeneous contribution $\chi_0$, provided that the matter is gathered around the mid-plane}. For $x \rightarrow 0$, we get from Eqs. \ref{eq:chi0_approx} and \ref{eq:chiq_approx}
\begin{flalign}
\label{eq:deltachi0}
\delta \chi_0 &= - \frac{\so}{\st} \sum_{n=0}^\infty{\frac{c_n}{2n+1}\left\{\frac{2n}{2n+1} \right.}\\
\nonumber
& \qquad  \qquad +  \left. \left[(-1)^n x^{2n} -1 \right] \left( \frac{\pi}{2}|x| - x\,\atan x \right) \right\}.
\end{flalign}
 We see that $\chi$ reaches a maximum value at $x=0$. This value is
\begin{flalign}
\chi_{\rm max} &= \chi_{0, \rm max} - \frac{\so}{\st} \sum_{n=1}^\infty{\frac{2n c_n}{(2n+1)^2}}.
\label{eq:gen_chimax}
\end{flalign}

\section{Application to softened gravity : a prescription for the softening length}
\label{sec:sg}

\subsection{The avoidance of point mass singularities}

As mentioned, the complete elliptic integral $\elik$ exhibits a logarithmic divergence as its modulus $k \rightarrow 1$. This is the direct consequence of the point mass singularity (i.e., $|\vec{r}-\vec{r'}| \rightarrow 0$). The determination of accurate potentials from Eqs. \ref{eq:psit} and \ref{eq:psif} is therefore not straightforward and requires a careful treatment of improper integrals \citep[e.g.,][]{syc90,hure05,hurepierens05}. This technical difficulty is usually circumvented by changing the relative separation according to
\begin{equation}
|\vec{r}-\vec{r'}| \quad \leftarrow \quad \sqrt{|\vec{r}-\vec{r'}|^2 + \lambda^2},
\label{eq:sl}
\end{equation}
where $\lambda \ne 0$ is a constant known as the ``softening length''. The main drawback is that softened gravity modifies Newton's law for gravitation both on short and long ranges. It lowers the magnitude of forces, enhances stability, and introduces a bias in models that is not easy to measure and interpret \citep[for stellar and gas discs see, e.g.,][]{paplin89,sayo90,romeo94,sommer98,dehnen01}.

In the case of gaseous discs of interest here, the modification of the relative distances according to Eq. \ref{eq:sl} changes the expressions for $\psif$ and $\psit$. It is however easy to show that the associated ``softened'' potentials denoted $\psif_{\rm s}$ and $\psit_{\rm s}$, respectively, can still be written in terms of Eqs. \ref{eq:psit} and \ref{eq:psif}, respectively, provided that the modulus $k$ is replaced by
\begin{equation}
\ks = \frac{2\sqrt{aR}}{\sqrt{(a+R)^2+z^2 + \lambda^2}},
\label{eq:ksoft}
\end{equation}
and $m$ is replaced by
\begin{equation}
\ms = \frac{2\sqrt{aR}}{\sqrt{(a+R)^2 + \lambda^2}}.
\label{eq:msoft}
\end{equation}
In disc models and simulations, one never (or rarely) computes $\psit$, or its fully asymmetric/tri-dimensional version \citep[see however][]{li09}. Instead, one computes $\psif$ in the framework of softened gravity, that is $\psif_{\rm s}$. The softening length $\lambda$ appearing in Eq. \ref{eq:msoft} must therefore be prescribed. We note that solving Eq. \ref{eq:msoft} for $\lambda$ leads to
\begin{flalign}
\frac{\lambda}{h} 
& =  \sqrt{\frac{{\ms'}^2}{1- {\ms'}^2} \frac{4aR}{h^2} - \frac{(a-R)^2}{h^2}},
\label{eq:slq}
\end{flalign}
where $\ms' = \sqrt{1-\ms^2}$ is the complementary modulus.

\begin{table}
\begin{tabular}{ll}
Reference & softening length $\lambda$ \\ \hline \\
\cite{paplin89} & $0.056 \times (\aout-\ain)\times f(a,R,\ain,\aout)$ \\
\cite{ars89} & $0.1\times R$ \\
\cite{sayo90} & $\{0.01,0.02\} \times \aout$  \\
\cite{shu90} & $0.1 \times R$ \\
\cite{mosa94} & $0.02 \times \aout$ \\
\cite{sterzik95}$^a$ & $0.06 \times \lambda_{\rm c}$  \\
\cite{lka97} & $0.1 \times \ain$, $0.1 \times \aout$, and $0.001 \times R$ \\
\cite{lka98} & $0.0001+0.01 R \times f(a,\ain,\aout)$ \\
\cite{tremaine01} & $\beta \times R$, with $\beta \approx 10^{-4} - 0.2$ \\
\cite{caunt01}$^b$ & $H=2h$ \\
\cite{baruteaumasset08} & $0.3 - 0.5 h$ (depending on scale height)  \\
\cite{li09}$^c$ & $\approx 0.17$ to $0.33 \times \Delta a$\\ \\
this work$^d$ & $f\left(\frac{h}{a},\frac{R}{a}\right)$\\
& $\approx h/e$ at $R=a$, homogeneous case; \\ 
& see Eqs. \ref{eq:chispecq_explicit}, \ref{eq:gen_chi}, \ref{eq:root} and \ref{eq:slq}\\ \\ \hline
\end{tabular}
\\ \\
$^a$$\lambda_{\rm c}$ is the critical wave length of disturbances.\\
$^b$concerns the magnetic potential.\\
$^c$$\Delta a$ is the grid spacing, 3D-disc.\\
$^d$axisymmetric limit, finite size disc (inner edge $\ain$, outer edge $\aout$), symmetry with respect to the mid-plane, finite size layer (thickness $2h=H$), explicit function of vertical stractification, local validity.\\
\caption{Various prescriptions for the softening length adopted in some simulations of self-gravitating gaseous discs (see also Fig. \ref{fig:sys.xfig.eps}).}
\label{tab:variouspresc}
\end{table}

\begin{figure}
\includegraphics[width=8.9cm, bb=32 51 706 522, clip=]{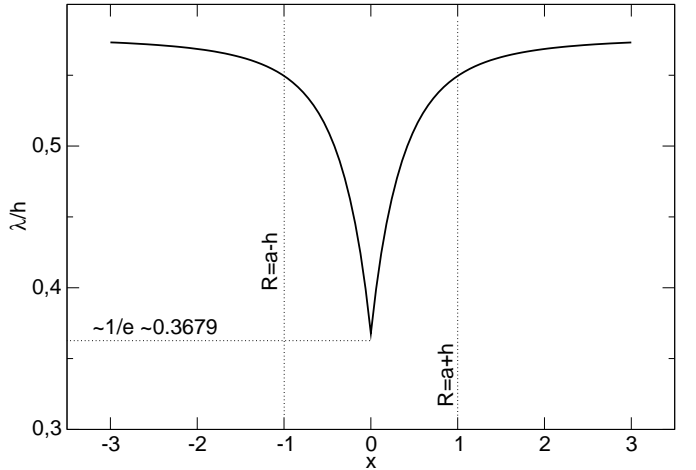}
\caption{Softening length normalized to the local semi-thickness $h$ versus $x$ as computed from Eq. \ref{eq:slq} for $h/a=0.1$ in the homogeneous case.}
\label{fig:lh0.eps}
\end{figure}

\subsection{A prescription for the softening length $\lambda$}
\label{subsec:lambda}

Many prescriptions for $\lambda$ have been proposed. In general, this is not a constant but a certain function of the radius and/or disc parameters. Table \ref{tab:variouspresc} gathers a few formulae for $\lambda$ used by different authors over twenty years. Although not exhaustive, this list clearly shows that there is no trend in magnitude and variation in space (and possible dependency on the disc parameters). The results obtained in Sect. \ref{sec:chif} can help substantially to define the appropriate prescription for $\lambda$. We can see from Eqs. \ref{eq:psif} and \ref{eq:psitbis} that the {\it gravitational potential in the mid-plane of a thin disc is equivalent to the softened potential caused by a flat disc} provided that $\lambda$ is the root of the equation
\begin{equation}
\st \ms \elik(\ms) -\int_{-h}^{+h}{\rho k \elik(k) dz} =0,
\label{eq:mps}
\end{equation} 
for all $R$. Only a numerical approach can yield the exact value of $\lambda$, if it exists. However, a good approximation to this root can be obtained by considering the expansion of the complete elliptic integral of the first kind over its complementary modulus, as considered in Sect \ref{sec:chif}. To the lowest order, Eq. \ref{eq:mps} becomes
\begin{equation}
\ln \frac{4}{\ms'} - \chi =0,
\end{equation}
where $\chi$ is given by Eq. \ref{eq:gen_chi}. In other words, this is
\begin{equation}
\ms'=4 e^{-\chi}.
\label{eq:root}
\end{equation}
We therefore conclude that the {\it appropriate prescription for $\lambda$, valid over a few disc thicknesses around the singularity $R=a$}, is given by Eq. \ref{eq:slq}, where $\ms'$ is found from Eq. \ref{eq:root}.

\subsection{Results for various stratifications}

It follows that $\lambda$ is proportional to $h$, and depends both on $x$ and $\epsilon$. It is also sensitive to the vertical stratification through the function $\chi$. Figure \ref{fig:lh0.eps} displays $\lambda/h$ versus $x$ for $h/a=0.1$ in the homogenous case (i.e., for $\chi \equiv \chi_0$, or $\delta \chi_0=0$). We see that, in the range of validity, $\lambda$ goes through a minimum for $x=0$. There, we have $k'_\pm \approx h/2a$ and $f_0 \approx 1/e$ and then the minimum is
\begin{equation}
\frac{\lambda}{h} \approx \frac{1}{e} \times \frac{1}{\sqrt{1 - \left(\frac{h}{2 ae}\right)^2}} \approx \frac{1}{e} \times \left[ 1 + \left(\frac{\epsilon}{\sqrt{2}e}\right)^2 \right].
\label{eq:sl_x=0_homo}
\end{equation}

\begin{figure}
\includegraphics[width=8.9cm, bb=-2 44 706 532, clip=]{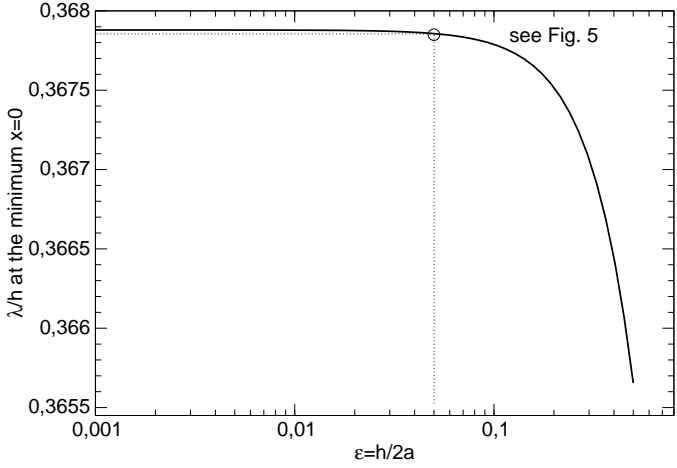}
\caption{The softening length at the minimum $x=0$ for the vertically homogeneous profile.}
\label{fig:lhvsaspectr.eps}
\end{figure}

This value is almost insensitive to the disc aspect ratio, provided that the disc is geometrically thin. This is shown in Figure \ref{fig:lhvsaspectr.eps}, where we plot $\lambda/h$ at the minimum versus $\epsilon$. We note however that the main variations of $\lambda$ {\it occur over a radial range of the order of $2h$} (i.e., the total disc thickness). The thinner the disc, the sharper the variation around the singularity.

\begin{figure}
\includegraphics[width=8.9cm, bb=0 51 706 522, clip=]{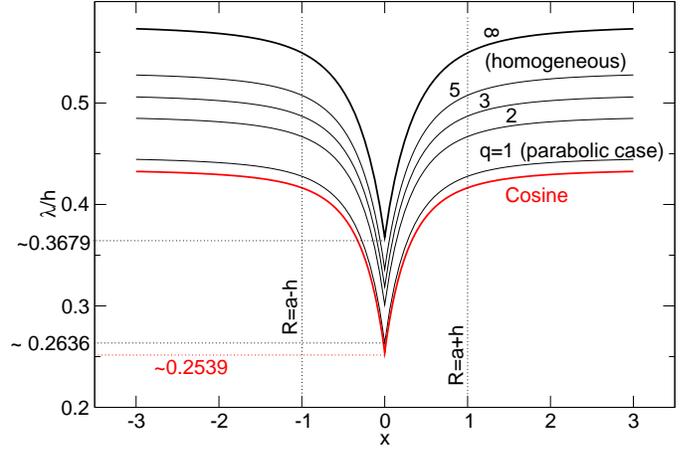}
\caption{Softening length normalized to the local semi-thickness $h$ versus $x$ as computed from Eq. \ref{eq:slq} for $h/a=0.1$. The density profile is according to Eq. \ref{eq:vertprofile} with $q=\{1,2,3,5, \infty\}$. The cosine profile is also shown.}
\label{fig:lh.eps}
\end{figure}

Figure \ref{fig:lh.eps} displays $\lambda/h$ for the mass density profile given by Eq. \ref{eq:vertprofile} and for $q=\{1, 2, 3, 5, \infty\}$. At $x=0$, we have the general formula
\begin{equation}
\frac{\lambda}{h} \approx \frac{e^{-(1 + \delta \chi_0)}}{\sqrt{1 - \left[\frac{h}{2 a}e^{- (1+\delta \chi_0)}\right]^2}}.
\label{eq:sl_x=0}
\end{equation}
Since $\delta \chi_0=\frac{1}{2q+1}$ in this case, we also have
\begin{equation}
\frac{\lambda}{h} 
 \approx e^{-\frac{2q+2}{2q+1}} =  \frac{1}{e} \times e^{-\frac{1}{2q+1}}.
\label{eq:sl_x=0_vertprofile}
\end{equation}

\subsection{An example of vertical mass density profile expandable in infinite series}

To illustrate the generality and power of the result, we consider a cosine profile of the form $\rho(z)=\cos \frac{\pi z}{2 h}$ which is a typical example where matter distribution is expandable in infinite series of the altitude. Actually, this profile can also be expanded by means of Eq. \ref{eq:genprofile} with the following coefficients :
\begin{equation}
c_n=\frac{(-1)^n}{(2n)!} \left(\frac{\pi}{2}\right)^{2n}.
\end{equation}

The corresponding function $\chi$ is then deduced from Eq. \ref{eq:gen_chi} and $\lambda$ is computed from Eqs. \ref{eq:slq} and \ref{eq:root}. Results are displayed in Figs. \ref{fig:chi.eps} and \ref{fig:lh.eps} for $\chi$ and $\lambda/h$ respectively. In particular, at $x=0$, we have
\begin{equation}
\delta \chi_0= - \pi \sum_1^\infty{\frac{(-1)^q q}{(2q+1)^2 (2q)!} \left(\frac{\pi}{2}\right)^{2q}} 
 = 0.371...
\end{equation}
which results in $\frac{\lambda}{h} \approx  e^{-1} \times 0.690 \approx 0.2539$ (this value would be obtained from Eq. \ref{eq:sl_x=0_vertprofile} for $q \approx 0.849$, which is close to the parabolic case).

\section{Concluding remarks}
\label{sec:conclusion}

In this paper, we have reported the first reliable prescription for the softening length $\lambda$ to be used in the numerical determination of the gravitational potential of geometrically thin discs within the framework of softened gravity. This expression has been found by rigorously comparing the Newtonian potential of a geometrically thin disc (of finite thickness) with the softened potential of a flat disc. This is a function of the radius and disc local aspect ratio, and also depends on vertical stratification. It is accurate at the singularity and around (typically a few disc thicknesses in radius). Although this formula is valid only locally, and obtained in the axisymmetric limit, it should help to improve the quality (realism, accuracy, and computing time) of 2D- and 3D-simulations\footnote{A Fortran 90 package called {\tt SingLe} is available at :\\ {\tt http://www.obs.u-bordeaux1.fr/radio/JMHure/intro2single.php}}. If necessary, the accuracy of the prescription can be improved by considering higher order terms in the expanded complete elliptic integrals of the first kind.

Finally, it would then be interesting to generalize our results i) to mass density profiles that extend to infinity, such as the Gaussian profile (which corresponds to vertically isothermal discs), and ii) to the entire disc since this prescription is expected to be inaccurate far from the singularity.

\begin{acknowledgements}
It is a pleasure to thank C. Baruteau, D. Bernard, A. Collioud, F. Hersant and A. Romeo for valuable comments.
\end{acknowledgements}

\bibliographystyle{aa}

\end{document}